\documentclass{article}
\usepackage{graphicx}
\usepackage{geometry}
\begin{document}

\title{Universal Power Law Scaling Near the Turning Points}
\author{M. Ali Saif\\
Department of Physics,\\
Faculty of Education,
University of Amran,
Amran,Yemen.\\
masali73@gmail.com}

\maketitle
\begin{abstract}

We show analytically and numerically that, the velocity $v_\pm$ of a particle near the turning points $x_0$ vanishes, i. e. $v_\pm\rightarrow 0$ as $x\rightarrow x_0$, according to the power law scaling $\left|v_\pm\right| \propto \left|x_0-x\right|^{\beta}$, where the exponent $\beta=1/2$ is independent of the particle mass and the force acting on it. We also show that, the time spends it any particle at each small interval $dx$ near the turning points diverges as $\tau\propto \left|x_0-x\right|^{\nu}$, with the exponent $\nu=-1/2$. Behavior we find here is very similar to power law scaling that had been found near the critical points for systems which undergo a phase transition.
\end{abstract}

Total mechanical energy $E$ for any particle in the absent of nonconservative force has a constant value given by \cite{gol}
\begin{eqnarray}
E=T(x)+V(x)
\end{eqnarray}

Where $T(x)$ is the kinetic energy function of the particle and $V(x)$ is the potential energy function. Now, if  $V(x)$ rises higher than the particle's total mechanical energy on either side, then the particle is stuck in the potential well Fig. 1. It rocks back and forth between the turning points.

\begin{figure}[htb]
 \includegraphics[width=70mm,height=60mm]{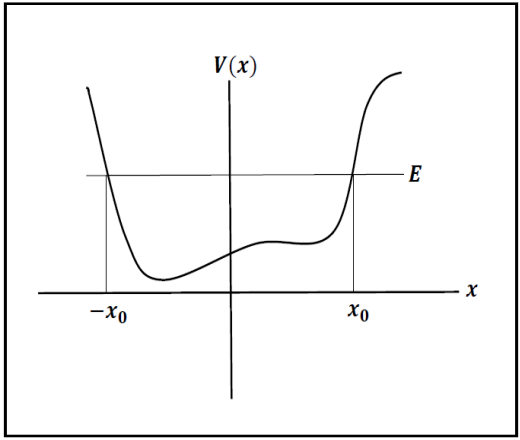}
 \caption{Turning points total mechanical energy representation for a particle moves in a potential well}
 \end{figure}

The velocity of such that particle between the turning points can be obtained from Eq. (1) to be as follows
\begin{eqnarray}
v_\pm(x)=\pm \sqrt{\frac{2}{m}(E-V(x))}
\end{eqnarray} 

Where $T(x)=\frac{1}{2}m v^2(x)$ and $m$ is the mass of the particle.

Here, we are going to study the velocity of such that particle near the turning points. And as we can see from Fig. 1, we can expand the potential energy function $V(x)$ in a Taylor series about the turning point $x_0$ to be:
\begin{eqnarray}
V(x)=V(x_0)+\grave{V(x_0)}(x-x_0)+\frac{1}{2}\grave{\grave{V(x_0)}}(x-x_0)^2+...
\end{eqnarray}

Near the turning points, where the value of $(x-x_0)$ is very small, we can ignore the higher order terms of the previous equation. Hence, substituting the value of $V(x)$ from Eq. (3) into Eq. (2), we get the following
\begin{eqnarray}
v_\pm(x)\simeq\sqrt{\frac{2}{m}(E-\left[V(x_0)+\grave{V(x_0)}(x-x_0)\right])}
\end{eqnarray}

However, at the turning points, we have $E=V(x_0)$, so that Eq. (4) becomes

\begin{eqnarray}
v_\pm(x)\simeq\pm\sqrt{\frac{2}{m}\grave{V(x_0)}(x_0-x)}
\end{eqnarray}
 
In the previous equation we have $\grave{V(x_0)}=-F(x_0)$, where $F(x_0)$ is the force acting on the particle at the turning point. Therefore, we can rewrite Eq. (5) as follows
 
 \begin{eqnarray}
v_\pm(x)\simeq\pm\sqrt{\frac{-2F(x_0)}{m}} \left[x_0-x\right]^{1/2}
\end{eqnarray} 

Previous equation describes the velocity of any particle near the turning points. Its clear that, near the turning points, $v_\pm\rightarrow 0$ as $x\rightarrow x_0$. That is, the velocity vanishes at turning points according to the power law $\left|v_\pm\right| \propto \left|x_0-x\right|^{1/2}$. This behavior is very similar to the behavior of systems those undergo a phase transition near the critical points \cite{sta}. Or even the behavior of dynamical systems those are close to a saddle-node bifurcation \cite{ste}.


In additional to that, the time spends it any particle at each small interval $dx$ as that particle approaches the turning points diverges as the following

\begin{eqnarray}
\tau\simeq(\frac{-2F(x_0)}{m})^{-1/2}\frac{1}{(x_0-x)^{1/2}} 
\end{eqnarray}

In the previous equation we used the fact that, $dt=\frac{dx}{v}$.
 
Hence, according to Eq. (7) the particle time diverges as the power law $\tau\propto \left|x_0-x\right|^{-1/2}$ near the turning points.

{\bf The Harmonic Oscillator}: The potential energy for a classical harmonic oscillator of a mass $m$ when that oscillator is attached to a spring of force constant $k$, is
\begin{eqnarray}
V(x)=\frac{1}{2}k x^2
\end{eqnarray}

So, from Eq. 2 the velocity of that oscillator will be 
\begin{eqnarray}
v_\pm(x)=\pm\sqrt{\frac{2E}{m}-\frac{kx^2}{m}}
\end{eqnarray}

Or
\begin{eqnarray}
v_\pm(x)=\pm\sqrt{\frac{k}{m}(\frac{2E}{k}-x^2)}
\end{eqnarray}

The turning points for the harmonic oscillator can be obtained from Eq. (10) to be

\begin{eqnarray}
x_0=\pm\sqrt{\frac{2E}{k}}
\end{eqnarray}

So that, we can rewrite Eq. (10) as follows 

\begin{eqnarray}
v_\pm(x)=\pm\sqrt{\frac{k}{m}(x_0^2-x^2)}=\pm \sqrt{\frac{k}{m}(x_0+x)(x_0-x)}
\end{eqnarray}

Near the turning point where $x\rightarrow x_0$, velocity of the harmonic oscillator can be given from Eq. (12) to be

\begin{eqnarray}
v_\pm(x)\simeq\pm \sqrt{\frac{k}{m}(2x_0)(x_0-x)}
\end{eqnarray}

Or
\begin{eqnarray}  
v_\pm(x)\simeq\pm \sqrt{\frac{2x_0k}{m}}\left[x_0-x\right]^{1/2}
\end{eqnarray}

Equation (14) is very similar to the general Eq. 6. and describes the velocity of the harmonic oscillator near the turning points. Its clear that, near the turning points, $v_\pm\rightarrow 0$ as $x\rightarrow x_0$. That is, the velocity of the harmonic oscillator vanishes at turning point according to the power law $\left|v_\pm\right| \propto \left|x_0-x\right|^{\frac{1}{2}}$. The corresponding harmonic time diverges as $\tau\propto\left|x_0-x\right|^{-1/2}$ at $x_0$.

In Fig. 2 we use the Eq. (10) to obtain the velocity of the harmonic oscillator, when the values of $m=1$ and $k=1$. We plot the velocity of the harmonic oscillator as function of $(x_0-x)$ Fig. 2(a), at three values of mechanical energy $E=50$, $E=200$ and $E=450$. Fig. 2(b) shows the double-logarithmic representation for the data in Fig. 2(a). For the best fit we find that, the velocity near turning point follows a power law with exponent equal $0.49991$, that is exactly as Eq. (14) says.

\begin{figure}[htb]
 \includegraphics[width=70mm,height=60mm]{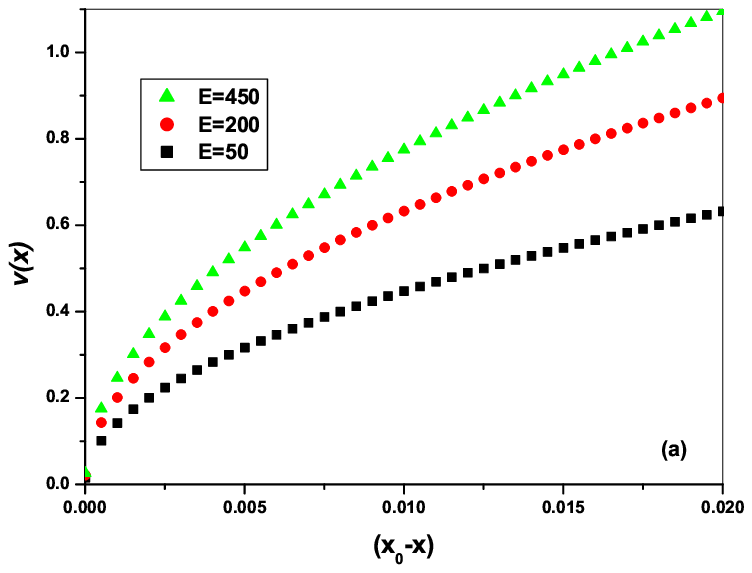}
 \includegraphics[width=70mm,height=60mm]{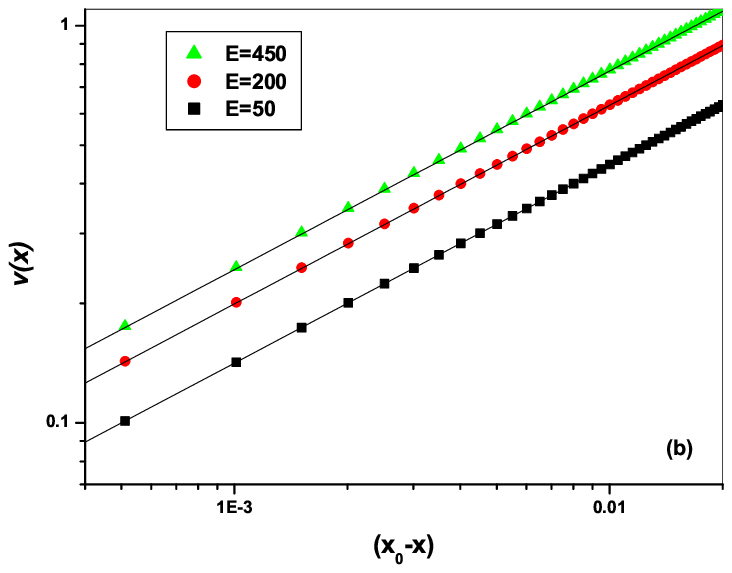}
 \caption{(a) Velocity of the harmonic oscillator near the turning point as function of $(x_0-x)$, when $m=1$ and $k=1$ at values of energy $E=50$, $E=100$ and $E=450$. (b) Double-logarithmic representation for the data in Fig. 2(a).}
 \end{figure}
 
 Fig. 3 shows the harmonic oscillator time as function of $(x_0-x)$ for the values of $m=1$, $k=1$ and $E=450$. Inset Fig. 3 shows double-logarithmic representation. Here, the power law fitting gives the exponent $-0.49996$ which coincides with the divergence Eq. (7).

\begin{figure}[htb]
 \includegraphics[width=70mm,height=60mm]{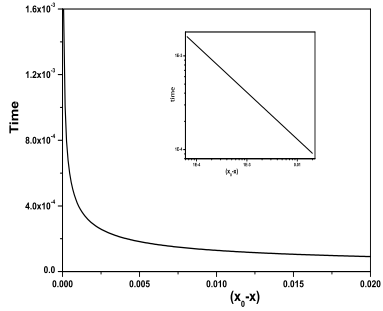}
\caption{Harmonic oscillator time near the turning point as function of $(x_0-x)$, when $m=1$ and $k=1$ at the value of energy $E=450$. Inset: Double-logarithmic representation.}
\end{figure}

So far from Eq. (14) we can prove that, the velocity near the turning points under an appropriate rescaling of energy $E\rightarrow\lambda E$, shows scaling behavior describes by the following relation:
\begin{eqnarray}
v(x,E)=\lambda^{-1/4} v(x,\lambda E)
\end{eqnarray}

In Fig. 4 we show that, the three energy curves in Fig. 2(a) collapse very well up to one curve using the previous scaling relation.  

\begin{figure}[htb]
 \includegraphics[width=70mm,height=60mm]{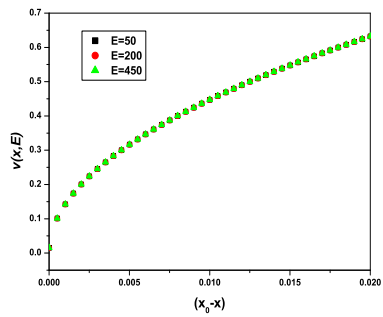}
 \caption{Three energy curves in Fig. 2(a) collapse very well up to one curve using the scaling relation Eq. (15).}
 \end{figure}
 
In conclusion we have shown that, the velocity of any particle near the turning points will vanish according to very well a power law scaling with an exponent equal to $1/2$. This behavior is similar to the behavior of systems which undergo a phase transition near the critical points. We also have shown that, the time spends it any particle at each small interval $dx$ as that particle approaches the turning points diverges with a power law with an exponent $-1/2$.

 \end{document}